\pdfoutput=1

\documentclass[%
 reprint,
superscriptaddress,
nofootinbib,
 amsmath,amssymb,
 aps,
]{revtex4-2}

\usepackage{graphicx}
\usepackage{dcolumn}
\usepackage{bm}
\usepackage[hidelinks]{hyperref}
\usepackage[mathlines]{lineno}
\usepackage{wasysym}
\usepackage{xcolor}
\usepackage{orcidlink}

\newcommand{\secref}{Sec.}
\newcommand{\figref}{Figure}
\newcommand{\tabref}{Table}
\usepackage[normalem]{ulem}
\newcommand\extrafootertext[1]{%
    \bgroup
    \renewcommand\thefootnote{\fnsymbol{footnote}}%
    \renewcommand\thempfootnote{\fnsymbol{mpfootnote}}%
    \footnotetext[0]{#1}%
    \egroup
}

\begin{document}

\preprint{APS/123-QED}

\title{Constraints on axion-like polarization oscillations in the cosmic microwave background with {\sc Polarbear}}

\author{Shunsuke Adachi~\orcidlink{0000-0002-0400-7555}}
 \affiliation{Hakubi Center for Advanced Research, Kyoto University, Kyoto 606-8501, Japan}
 \affiliation{Department of Physics, Faculty of Science, Kyoto University, Kyoto 606-8502, Japan}
 
\author{Tylor Adkins~\orcidlink{0000-0002-3850-9553}}
\affiliation{Department of Physics, University of California, Berkeley, Berkeley, CA 94720, USA}

\author{Kam Arnold}
\affiliation{Department of Physics, University of California, San Diego, La Jolla, CA 92093, USA}

\author{Carlo Baccigalupi~\orcidlink{0000-0002-8211-1630}}
\affiliation{Astrophysics and Cosmology, The International School for Advanced Studies (SISSA), Trieste 34136, Italy}
\affiliation{Astrophysics and Cosmology, Institute for Fundamental Physics of the Universe (IFPU), Trieste 34151, Italy}
\affiliation{Astrophysics and Cosmology, The National Institute for Nuclear Physics (INFN), Trieste 34127, Italy}

\author{Darcy Barron~\orcidlink{0000-0002-1623-5651}}
\affiliation{Department of Physics and Astronomy, University of New Mexico, Albuquerque, NM 87131, USA}

\author{Kolen Cheung~\orcidlink{0000-0002-7764-378X}}
\affiliation{Department of Physics, University of California, Berkeley, Berkeley, CA 94720, USA}
\affiliation{Space Sciences Laboratory, University of California, Berkeley, Berkeley, CA 94720, USA}
\affiliation{Computational Cosmology Center, Lawrence Berkeley National Laboratory, Berkeley, CA 94720, USA}

\author{Yuji Chinone~\orcidlink{0000-0002-3266-857X}}
\email[Corresponding authors: ]{jspisak@ucsd.edu, chinoney@gmail.com}
\affiliation{QUP (WPI), KEK, Tsukuba, Ibaraki 305-0801, Japan}
\affiliation{Kavli Institute for the Physics and Mathematics of the Universe (WPI), UTIAS, The University of Tokyo, Kashiwa, Chiba 277-8583, Japan}

\author{Kevin T. Crowley~\orcidlink{0000-0001-5068-1295}}
\affiliation{Department of Physics, University of California, San Diego, La Jolla, CA 92093, USA}

\author{Josquin Errard~\orcidlink{0000-0002-1419-0031}}
\affiliation{Universit\'{e} de Paris Cit\'e, CNRS, Astroparticule et Cosmologie, F-75013 Paris, France}

\author{Giulio Fabbian~\orcidlink{0000-0002-3255-4695}}
\affiliation{Center for Computational Astrophysics, Flatiron Institute, New York, NY 10003, USA}
\affiliation{School of Physics and Astronomy, Cardiff University, Cardiff CF24 3AA, UK}

\author{Chang Feng~\orcidlink{0000-0001-7438-5896}}
\affiliation{Department of Astronomy, School of Physical Sciences, University of Science and Technology of China, Hefei 230026, China}

\author{Raphael Flauger}
\affiliation{Department of Physics, University of California, San Diego, La Jolla, CA 92093, USA}

\author{Takuro Fujino~\orcidlink{0000-0002-1211-7850}}
\affiliation{Graduate School of Engineering Science, Yokohama National University, Yokohama 240-8501, Japan}

\author{Daniel Green~\orcidlink{0000-0001-5496-0347}}
\affiliation{Department of Physics, University of California, San Diego, La Jolla, CA 92093, USA}

\author{Masaya Hasegawa~\orcidlink{0000-0003-1443-1082}}
\affiliation{Institute of Particle and Nuclear Studies (IPNS), High Energy Accelerator Research Organization (KEK), Tsukuba 305-0801, Japan}
\affiliation{QUP (WPI), KEK, Tsukuba, Ibaraki 305-0801, Japan}
\affiliation{SOKENDAI, SOKENDAI, Tsukuba 305-0801, Japan}

\author{Masashi Hazumi~\orcidlink{0000-0001-6830-8309}}
\affiliation{QUP (WPI), KEK, Tsukuba, Ibaraki 305-0801, Japan}
\affiliation{Institute of Particle and Nuclear Studies (IPNS), High Energy Accelerator Research Organization (KEK), Tsukuba 305-0801, Japan}
\affiliation{Institute of Space and Astronautical Science (ISAS), Japan Aerospace Exploration Agency (JAXA), Sagamihara 252-5210, Japan}
\affiliation{Kavli Institute for the Physics and Mathematics of the Universe (WPI), UTIAS, The University of Tokyo, Kashiwa, Chiba 277-8583, Japan}

\author{Daisuke Kaneko~\orcidlink{0000-0003-3917-086X}}
\affiliation{QUP (WPI), KEK, Tsukuba, Ibaraki 305-0801, Japan}

\author{Nobuhiko Katayama}
\affiliation{Kavli Institute for the Physics and Mathematics of the Universe (WPI), UTIAS, The University of Tokyo, Kashiwa, Chiba 277-8583, Japan}

\author{Brian Keating~\orcidlink{0000-0003-3118-5514}}
\affiliation{Department of Physics, University of California, San Diego, La Jolla, CA 92093, USA}

\author{Akito Kusaka}
\affiliation{Physics Division, Lawrence Berkeley National Laboratory, Berkeley, CA 94720, USA}
\affiliation{Department of Physics, The University of Tokyo, Tokyo 113-0033, Japan}
\affiliation{Kavli Institute for the Physics and Mathematics of the Universe (WPI), UTIAS, The University of Tokyo, Kashiwa, Chiba 277-8583, Japan}
\affiliation{Research Center for the Early Universe, Graduate School of Science, The University of Tokyo, Tokyo 113-0033, Japan}

\author{Adrian T. Lee~\orcidlink{0000-0003-3106-3218}}
\affiliation{Department of Physics, University of California, Berkeley, Berkeley, CA 94720, USA}
\affiliation{Physics Division, Lawrence Berkeley National Laboratory, Berkeley, CA 94720, USA}

\author{Yuto Minami~\orcidlink{0000-0003-2176-8089}}
\affiliation{Research Center for Nuclear Physics, Osaka University, Ibaraki 567-0047, Japan}

\author{Haruki Nishino~\orcidlink{0000-0003-0738-3369}}
\affiliation{Research Center for the Early Universe, Graduate School of Science, The University of Tokyo, Tokyo 113-0033, Japan}

\author{Christian L. Reichardt~\orcidlink{0000-0003-2226-9169}}
\affiliation{School of Physics, The University of Melbourne, Parkville 3010, Australia}

\author{Praween Siritanasak~\orcidlink{0000-0001-6830-1537}}
\affiliation{National Astronomical Research Institute of Thailand, Chiangmai 50180, Thailand}

\author{Jacob Spisak~\orcidlink{0000-0003-1789-8550}}
\email[Corresponding authors: ]{jspisak@ucsd.edu, chinoney@gmail.com}
\affiliation{Department of Physics, University of California, San Diego, La Jolla, CA 92093, USA}

\author{Osamu Tajima~\orcidlink{0000-0003-2439-2611}}
\affiliation{Department of Physics, Faculty of Science, Kyoto University, Kyoto 606-8502, Japan}

\author{Satoru Takakura~\orcidlink{0000-0001-9461-7519}}
\affiliation{Department of Astrophysical and Planetary Sciences, University of Colorado, Boulder, CO 80309, USA}

\author{Sayuri Takatori~\orcidlink{0000-0002-8705-9624}}
\affiliation{Research Institute for Interdisciplinary Science (RIIS), Okayama University, Okayama 700-8530, Japan}

\author{Grant Paul Teply}
\affiliation{Department of Physics, University of California, San Diego, La Jolla, CA 92093, USA}

\author{Kyohei Yamada~\orcidlink{0000-0003-0221-2130}}
\affiliation{Department of Physics, The University of Tokyo, Tokyo 113-0033, Japan}


\collaboration{The {\sc Polarbear} Collaboration}

\date{June 15, 2023}

\begin{abstract}
Very light pseudoscalar fields, often referred to as axions, are compelling dark matter candidates and can potentially be detected through their coupling to the electromagnetic field. Recently a novel detection technique using the cosmic microwave background~(CMB) was proposed, which relies on the fact that the axion field oscillates at a frequency equal to its mass in appropriate units, leading to a time-dependent birefringence. For appropriate oscillation periods this allows the axion field at the telescope to be detected via the induced sinusoidal oscillation of the CMB linear polarization. We search for this effect in two years of {\sc Polarbear} data. We do not detect a signal, and place a median $95 \%$ upper limit of $0.65 ^\circ$ on the sinusoid amplitude for oscillation frequencies between $0.02\,\text{days}^{-1}$ and $0.45\,\text{days}^{-1}$, which corresponds to axion masses between $9.6 \times 10^{-22} \, \text{eV}$ and $2.2\times 10^{-20} \,\text{eV}$. Under the assumptions that 1) the axion constitutes all the dark matter and 2) the axion field amplitude is a Rayleigh-distributed stochastic variable, this translates to a limit on the axion-photon coupling $g_{\phi \gamma} < 2.4 \times 10^{-11} \,\text{GeV}^{-1} \times ({m_\phi}/{10^{-21} \, \text{eV}})$.
\end{abstract}

\maketitle

\tableofcontents

\section{Introduction} \label{sec:intro}
The nature of dark matter, particularly its non-gravitational interactions, remains one of the biggest open questions in cosmology and particle physics. One possibility that has recently received significant attention is low-mass bosonic dark matter \cite{Duffy_2009, Marsh_2016, Hui_2021}. The canonical example is the original QCD axion, a pseudo-Nambu Goldstone boson associated with the spontaneous breaking of a U(1) symmetry proposed to solve the strong CP problem. \cite{Peccei_1977, Pecci_1977_2, Weinberg_1978, Wilczek_1978}. More generally, a broader class of pseudoscalar fields with small masses and couplings to the standard model have been considered. These are often called axion-like-particles and are not necessarily solutions to the strong CP problem; nevertheless, we will refer to them as axions in this work for brevity. String theory generically predicts the existence of many such axions populating a wide range of masses and couplings, which is sometimes called the axiverse \cite{Arvanitaki_2010, Svrcek_2006}. Axions with astrophysically large de Broglie wavelengths ($\lambda \sim 1 \,\text{kpc}$ for axion mass $m_\phi \sim 10^{-22} \,\text{eV}$) are a particularly intriguing dark matter candidate because of their ability to act as fuzzy dark matter, which can potentially resolve conflicts between small-scale predictions of cold dark matter models and observations~\cite{Hui_2017, Hu_2000}. 

One way to detect axions is via their interaction with electromagnetism. Laboratory experiments such as ADMX \cite{Caldwell_2017} and  ABRACADABRA \cite{Kahn_2016}, for example, exploit the coupling between axions and magnetic fields to set limits on the axion dark matter in the QCD mass range. Reference \cite{Irastorza_2018} provides an overview of experimental approaches. It is well known that the coupling between the electromagnetic field and a pseudoscalar field generates an effective birefringence, rotating linearly polarized light \cite{HARARI_1992, Carroll_1998}.  The axion is well modelled as a classical field, which, when the axion mass is less than the Hubble rate, oscillates at a frequency equal to its mass. Fedderke et. al. \cite{Fedderke_2019}, hereafter F19, pointed out two novel effects in the cosmic microwave background~(CMB) caused by this oscillation and birefringence. The first is the suppression of the overall CMB polarization signal due to averaging over many axion oscillation periods during recombination, which was constrained in F19. The second is a coherent, all-sky oscillation of the CMB's linear polarization due to the oscillation of the axion field \textit{at the telescope}. This effect can be constrained when the oscillation period is appropriate for CMB experiments, e.g. hours to years, corresponding to masses in the $10^{-19} \,\text{eV} - 10^{-22} \,\text{eV}$ range. This also happens to be the mass range in which the axion can act as fuzzy dark matter \cite{Hui_2017}. In this work, we search for this effect using data from the {\sc Polarbear} experiment. 

This signal has been constrained by other CMB experiments: BICEP/Keck, in \cite{BK2020, BK_2022} (the latter hereafter BK22), and the South Pole Telescope~(SPT) in \cite{SPT_2022} (hereafter SPT22). Our analysis is similar to these analyses, with two primary differences. The first is that we estimate the CMB polarization angle using $C_\ell^{EB}$ power spectra rather than in $Q/U$ pixel space. This allows us to use the theoretical $C_\ell^{EE}$ power spectrum from precisely measured cosmological parameters as a polarization template rather than coadded $Q/U$ maps, and also facilitates easier systematic error checks. The second is that we model the amplitude of the axion field at the telescope as a stochastic, Rayleigh-distributed variable rather than assuming a fixed value corresponding to the mean Milky-Way halo density \cite{Foster_2018}. The need for this approach was pointed out in \cite{Centers_2021} and weakens the median constraint on the axion-photon coupling constant by a factor of $2.2$ in our analysis.  

The birefringence signature generated by axion dark matter can be constrained by other astrophysical polarized sources, and many other authors have used this effect to place constraints in a similar mass range using pulsars, active galactic nuclei, protoplanetary disks, Sagittarius~A$^*$, and black hole superradiance \cite{Fujita_2019, Caputo_2019, Liu_2020, Ivanov_2019, Basu:2020gsy, Yuan:2020xui, Chen_2021, Castillo_2022}. The CMB, however, has several attractive features that make it ideal for this type of analysis. The signal is entirely due to the axion field at the telescope, so we do not require any modelling of the field at the source (during the release of the CMB). The template polarization signal from the CMB has minimal time-dependent contamination and is extremely well measured across many experiments. Finally, {\sc Polarbear} and other CMB instruments have a long history of precision CMB polarization measurements with well-understood noise properties. These factors all serve to mitigate the systematic error in this analysis. 

In addition to birefringence-based measurements, bounds from cosmological structure like the Lyman-$\alpha$ forest and Milky Way satellites have placed constraints on the minimum allowed axion mass \cite{Ir_i__2017, Rogers_2021, DES:2020fxi} if it is fuzzy dark matter. A constraint on the minimum allowed mass has also been derived from the impact of dynamical heating on the velocity dispersions of ultra-faint dwarf galaxies \cite{dalal_2022}. These bounds strongly constrain the allowed parameter space for fuzzy dark matter, but are subject to different systematic and modelling uncertainties than birefringence analyses. Upper bounds on the axion-photon coupling which are constant in the mass range we consider have been derived from axion-photon conversion in the Sun, supernova 1987A, quasar H1821+643 and in the intracluster medium \cite{CAST_2017, Payez_2015, 10.1093/mnras/stab3464, Buen-Abad:2020zbd}.

To perform this analysis we use the first two seasons of data from the {\sc Polarbear} experiment, which measured the polarization of the CMB during 2012--2016. During the first two seasons of observations, {\sc Polarbear} observed three small sky patches in order to measure gravitationally lensed $B$-modes. These results were presented in \cite{PB_2014} (hereafter PB14) and \cite{PB_2017} (hereafter PB17). The telescope has an angular resolution of 3.5~arcmin and reported measurements of the $B$-mode power spectrum up to multipoles of $\ell=2100$ from the first two seasons, including the angular scales at which the CMB polarization signal is the strongest. {\sc Polarbear} measurements have previously been used to constrain anisotropic birefringence and primordial magnetic fields \cite{PB_2015}: this work is entirely seperate, however, because we are searching for \textit{time-dependent} oscillations of the isotropic birefringence angle, whereas \cite{PB_2015} considered time-independent birefringence.

The rest of the paper is organized as follows. In \secref~\ref{sec:telescope}, we describe the {\sc Polarbear} instrument and observations. In \secref~\ref{sec:analysis}, we detail the analysis procedure used to generate CMB polarization angles and search for an axion signal. In \secref~\ref{sec:null_tests} and \ref{sec:systematics}, we describe the null tests and systematics estimates used to validate the dataset. The results are presented in \secref~\ref{sec:results}, and the conclusion in \secref~\ref{sec:conclusion}.

\section{First and Second Season Observations of the {\sc Polarbear} Instrument} \label{sec:telescope}
The {\sc Polarbear} experiment consisted of a cryogenic receiver attached to a two-mirror reflective telescope, the 2.5~m Huan Tran Telescope. It was located at the James Ax Observatory in the Atacama Desert in Chile at an elevation of 5{,}190~m.
The receiver contained 1,274 transition-edge sensors arranged into 637 polarization-sensitive pixels situated in 7 detector wafers on the focal plane,
which was cooled to 0.3~K, and observed at a single frequency centered at 150~GHz. We will report data from the first and second season observations, which occurred from May 2012 to June 2013 and October 2013 to April 2014, respectively. More details about the {\sc Polarbear} receiver and telescope can be found in \cite{Arnold_2012} and \cite{Kermish_2012}.

The {\sc Polarbear} observing strategy during the first and second seasons is described in detail in PB14 and PB17, and will be summarized here. Three separate sky patches were observed, each with an effective sky area of $7\text{--}9$ square degrees. In this work, an ``observation'' will refer to a single, continuous measurement of one patch which lasts until the patch is no longer visible, typically 4--8 hours. Each observation consists of many 15-minute constant elevation scans~(CESs). During one CES, the telescope scans back and forth in azimuth repeatedly at a constant elevation, and as the sky rotates, the entire patch is observed. The telescope then changes elevation and repeats the process. A typical observing day involves sequentially observing all three patches. This observing strategy allows us to probe oscillation periods longer than about 2~days.

During the first two seasons the telescope observed with a cryogenic half-wave plate (HWP) located on the sky-side of the lenses. During the first season it was periodically rotated between observations in order to mitigate systematic errors, and during the second season it remained fixed. In PB17, it was noted that this generated some uncertainty in the absolute polarization angle of the instrument during the first season and between the two seasons. This source of systematic error limits the maximum oscillation period we can assess to 50~days, as discussed in \secref~\ref{sec:systematics}. 

\section{Analysis Method} \label{sec:analysis}

\subsection{Expected Signal}
Following F19, the axion-photon coupling in the Lagrangian can be written as
\begin{equation}
    \mathcal{L} = - \frac{1}{4} g_{\phi \gamma} \phi F_{\mu \nu} \tilde{F}^{\mu \nu} \text{,}
\end{equation}
where $\phi$ is the axion field, $g_{\phi \gamma}$ is the axion-photon coupling, $F_{\mu \nu}$ is the electromagnetic field tensor and $\tilde{F}^{\mu \nu}$ is its dual. Assuming that the amplitude is small enough so the potential is well-approximated by $V(\phi) = m_{\phi}^2 \phi^2/2$, the axion field at the telescope is well-described by $\phi(t) = \phi_0 \sin(m_\phi t + \theta)$. We treat the amplitude $\phi_0$ and phase $\theta$ as constants because the duration of the experiment is much less than the axion coherence time, as discussed in \secref~\ref{sec:constraints}. The polarization angle of the CMB due to the axion field at the telescope is then (F19)
\begin{equation}
    \beta_{\text{CMB}}(t) = \frac{g_{\phi \gamma} \phi_0}{2} \sin(m_\phi t  + \theta).
\end{equation}
This can be conveniently parameterized as
\begin{equation} \label{signal}
    \beta_{\text{CMB}}(t) = A \sin(2 \pi f t  + \theta) .
\end{equation}
This is a sinusoid with unknown amplitude $A$, frequency $f$, and phase $\theta$. The basic unit of time in the {\sc Polarbear} survey schedule is one observation of a single patch, which can last up to 8 hours. For each observation, therefore, we construct CMB maps and estimate a single angle, generating several hundred angles over the course of the two years of data. We will then form a likelihood to search for the presence of a sinusoidal signal in this data. 

\subsection{Angle Estimation Procedure}
There are multiple ways to estimate a rotation angle from the observation maps. The most direct way is to search for the rotation in the Stokes $Q$ and $U$ parameter maps by comparing them to a template of $Q$ and $U$ maps containing only the unrotated CMB. This is the method employed in BK22 and SPT22. Another way is to transform the $Q$ and $U$ maps into $E$-mode and $B$-mode maps, then use a single-observation $C_\ell^{EB}$ spectrum to estimate a rotation angle. While the latter method has the disadvantage of requiring more computationally-intensive steps, we choose to implement it for the following reasons. The first is that we can use the extremely well-determined theoretical $C_\ell^{EE}$ from other experiments as our polarization template, rather than a $Q/U$ template map created by coadding our observations. While the $C_\ell^{EE}$ template is affected by sample variance, this effect is not important for the $\ell$ range we use. The second advantage is that the power spectra approach is very similar to the approach taken in the construction of the full, coadded spectra in PB17. This lets us re-use most elements of this validated pipeline, including many of the same systematics estimates.

The method of estimating a time-independent rotation angle using the $C_\ell^{EB}$ power spectrum is well established \cite{Keating_2012}, and has been used in many analyses, including PB17, to correct for an overall telescope miscalibration angle. The method we will employ to search for a \textit{time-varying} rotation angle is similar, except that we must construct a spectrum for every observation. This will be used to estimate a single rotation angle for each observation, constant over the duration of the observation. We can then search for a time-varying signal in the  timestream of these angles. For a single observation with CMB rotation angle $\alpha$, in the absence of noise and foreground contamination (which are addressed in \secref~\ref{sec:spectra_to_angles} and \secref~\ref{sec:foregrounds_other} respectively), the observed Fourier transformed $E/B$-mode coefficients are 

\begin{flalign}
    E^{\text{\text{obs}}}_{\ell m} &= \cos(2 \alpha) E^{\text{CMB}}_{\ell m} - \sin(2 \alpha) B^{\text{CMB}}_{\ell m} \\
    B^{\text{obs}}_{\ell m} &= \sin(2 \alpha) E^{\text{CMB}}_{\ell m} + \cos(2 \alpha) B^{\text{CMB}}_{\ell m}.
\end{flalign}

This analysis is sensitive to signals of amplitude ${\approx} 1^\circ$, so we use the small angle approximation

\begin{flalign}
    E^{\text{obs}}_{\ell m} &= E^{\text{CMB}}_{\ell m}  - 2 \alpha B^{\text{CMB}}_{\ell m} \\
    B^{\text{obs}}_{\ell m} &= 2 \alpha E^{\text{CMB}}_{\ell m} + B^{\text{CMB}}_{\ell m}. \label{eq:Bobs}
\end{flalign}

Since the CMB is $E$-mode dominated, to leading order all of the rotation information is contained in the $2 \alpha E^{\text{CMB}}_{\ell m}$ term in Eq.~\eqref{eq:Bobs}. In order to recover an angle from a single observation, then, we can construct a power spectrum using one $B$-mode map correlated with the full, coadded $E$-mode maps. The spectra for observation $j$, rotated by angle $\alpha_j$, is

\begin{flalign}
\begin{split}
    C_{\ell, j}^{EB, \text{obs}} &= \frac{1}{N} \frac{1}{(2 \ell+1)}\sum_{i}^N \sum_{m} E^{\text{obs}}_{\ell m, i} (B^{\text{obs}}_{\ell m, j})^* \\
    &= 2 \alpha_j C_\ell^{EE, \text{CMB}} .
\end{split}
\end{flalign}

The $E$-mode maps have been coadded over all observations $i = 1, \dotsc, N$. The intrinsic $C_\ell^{EB, \text{CMB}}$ has been set to zero, and we have neglected $C_\ell^{BB, \text{CMB}}$ because it is $\ll C_\ell^{EE, \text{CMB}}$.

\subsection{Mapmaking} \label{sec:mapmaking}
The mapmaking procedure is nearly identical to the one used in ``Pipeline A'' of PB17, and will be briefly reviewed here. To make maps of the polarized sky, the raw time-ordered data (TOD) undergo a series of quality cuts, are converted to CMB temperature units, and then differenced to form polarization timestreams. The timestreams are filtered to remove high and low frequency noise as well as scan-synchronous signals then combined with the detector pointing data to make maps. The maps are apodized with noise-weighted masks, which cover point sources as well. This procedure yields a single $Q$ and $U$ map for each observation. Finally, $Q$ and $U$ maps are transformed to $E$ and $B$ maps using the pure $B$-mode transform~\cite{Smith_2006}. Due to the filtering, they are biased estimates of the true sky signal.

The chief difference between the maps used in this analysis and those in PB17 is the apodization. PB17 uses a separate mask for each observation, whereas this analysis uses one apodization mask common to every observation within each patch. This is done in order to simplify the calculation of the mode-mixing matrix, discussed in \secref~\ref{sec:maps_to_spectra}. However, this masking procedure does sub-optimally weight each pixel within a given observation, leading to an increase in noise. In some observations the increase in noise from edge pixels is excessive, and they are removed from the analysis if the map noise increase exceeds a certain threshold. The same cut is applied to observations that have excessive noise on one split in a null test as well, as described in \secref~\ref{sec:null_analysis}. The overall impact of this apodization procedure is an effective noise increase of about $15 \%$.
Improving the apodization procedure to lessen this noise hit is an area of improvement for future analyses.

\subsection{Maps to Spectra} \label{sec:maps_to_spectra}
The procedure to construct single-observation $C_\ell^{EB}$ spectra as outlined above is identical to PB17, with the modification that only one $B$-mode map is used per observation. The co-added $E$-mode maps and a single $B$-mode map are used to construct pseudospectra ($\Tilde{C}_{\ell}$), which are biased estimates of the true spectra:
\begin{equation}
    \Tilde{C}_{\ell, j}^{EB} = \frac{1}{N_{\ell}} \sum_{k \in {\rm bin}_\ell} \Tilde{m}^{B*}_{jk} \left( \frac{1}{\sum_{i \neq j} w_i} \sum_{i \neq j} w_i \Tilde{m}^E_{ik} \right) .
\end{equation}
The $\Tilde{m}_{jk}$ denotes a Fourier-transformed map with 2D wavevector $k$ and observation $j$. The flat-sky approximation is used due to the size of the {\sc Polarbear} patches, and the spectra are combined into $\ell$ bins of width $\Delta \ell = 40$. $N_{\ell}$ is the number of wavevectors in bin $\ell$. The term in the parentheses represents the coadded $E$-mode map, with observation weights $w_i$. Observation $j$ is removed from the coadd to eliminate noise bias, since noise is assumed to be uncorrelated between observations.

The estimated true spectra ($\hat{C_b}$) are calculated from the pseudospectra using
\begin{equation}
    \hat{C}_{b,j}^{EB} \equiv \sum_{b'} K_{bb',j}^{-1} P_{b'\ell} \tilde{C}_{\ell,j}^{EB}
\end{equation}
and
\begin{equation}
    K_{bb',j} = \sum_{\ell\ell'} P_{b\ell} M_{\ell\ell', \text{patch}} F_{\ell', j} B_{\ell'}^2 Q_{\ell'b'} .
\end{equation}
All variables in this calculation have the same meaning as in PB17 but some are re-calculated for this analysis. The mode-mixing matrices $M_{\ell\ell', \text{patch}}$ correct for the effects of apodization and are calculated analytically for each patch using the patch apodization mask, discussed in \secref~\ref{sec:mapmaking}. The filter transfer functions $F_{\ell', j}$ correct for the effects of time-domain filters for observation $j$, $B_{\ell}$ corrects for the beam, and $P_{b\ell}$ and $Q_{\ell'b'}$ are binning matrices. There are four bins $b$ centered at $\ell = [700, 1100, 1500, 1900]$, each of width $\Delta \ell = 400$.

\subsection{Spectra to Angles} \label{sec:spectra_to_angles}
\begin{figure*}
    \includegraphics[height=0.24\textheight]{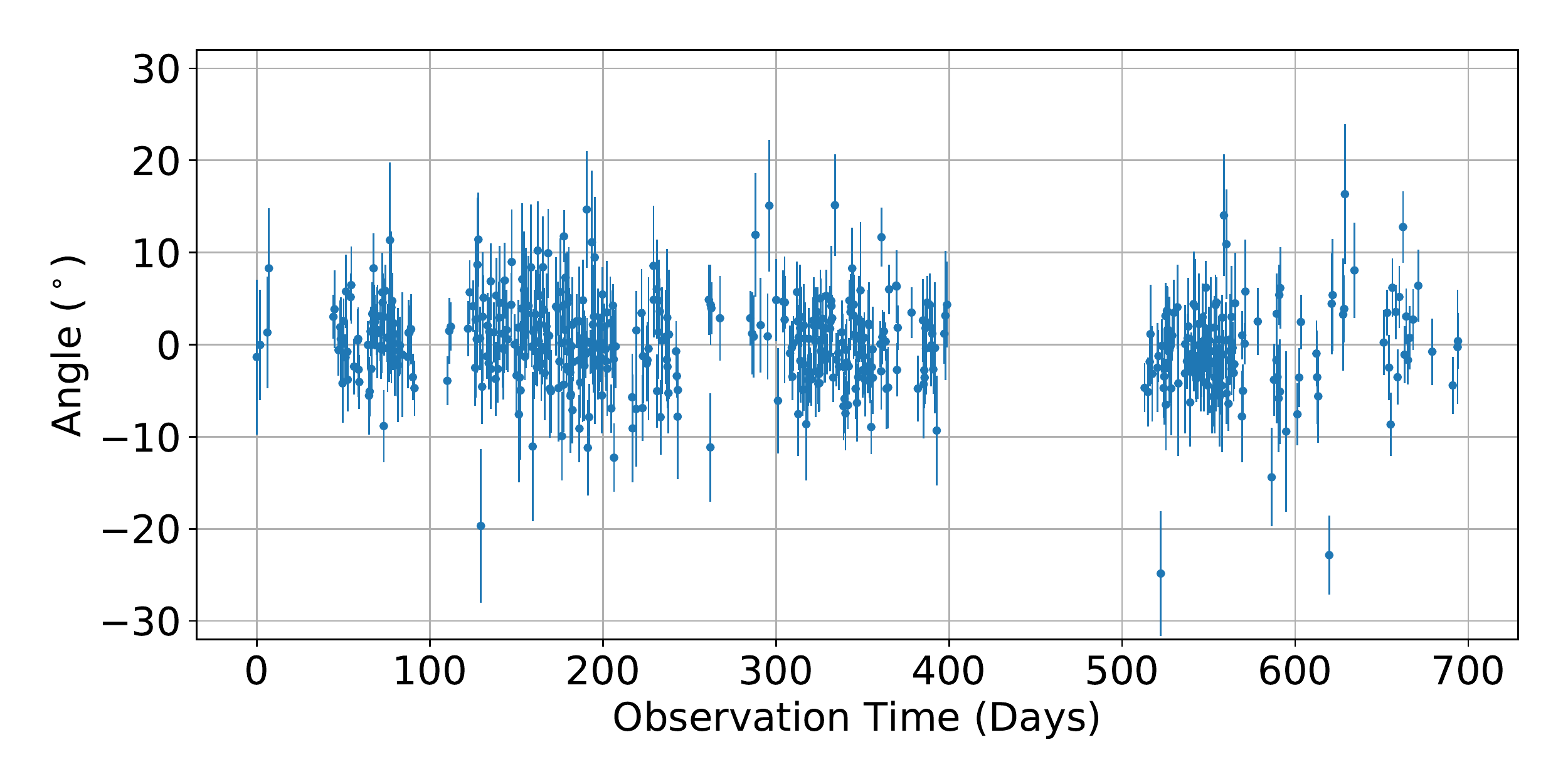}
    \includegraphics[height=0.24\textheight]{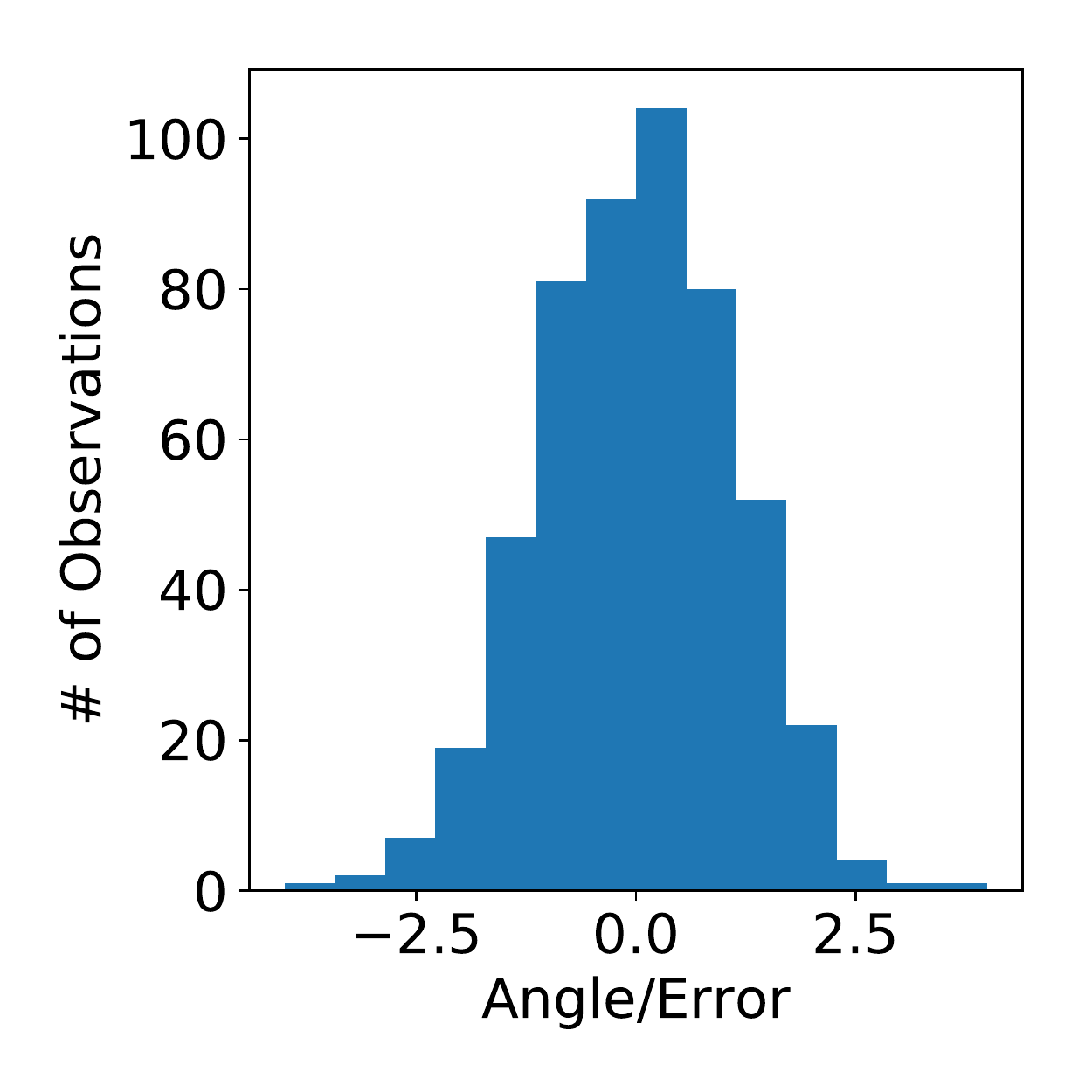}
    \caption{\textit{Left}: The 515 estimated per-observation rotation angles $\hat{\alpha}_j$ used in this analysis.
      \textit{Right}: A histogram of the angles normalized by their error, $\hat{\alpha}_j / \sigma_j$. The angles are Gaussian distributed.}
    \label{fig:angle_estimate}
\end{figure*}

Since the noise between different observations is uncorrelated, the noise between different $C_\ell^{EB}$ spectra is also uncorrelated. We will neglect the impact of foregrounds in the angle estimation: all three patches were chosen to have low foregrounds, and this systematic effect is discussed in \secref~\ref{sec:foregrounds_other}. The estimated rotation angle $\hat{\alpha}_j$ for observation $j$ is then obtained by minimizing
%
\begin{equation}
    \chi^2(\hat{\alpha}_j) = \sum_{b b'} (\hat{C}_{b,j}^{EB} - 2 \hat{\alpha_j} C_{b}^{EE, \text{th}}) (V^{EB}_{b b'})^{-1} (\hat{C}_{b'j}^{EB} - 2 \hat{\alpha_j} C_{b'}^{EE, \text{th}}) .
\end{equation}
The theoretical power spectrum $C_{b'}^{EE, \text{th}}$ is calculated from the WMAP 9-year data. We use WMAP rather than Planck data because there are negligible differences for the purpose of our analysis and the WMAP results are already integrated in our pipeline. The covariance matrix $V^{EB}_{b b'}$ is calculated from 500 simulated maps that contain $\Lambda$CDM signal and instrumental noise. These simulated maps are generated by constructing maps containing only $\Lambda$CDM signal and scanning them into TOD. White noise from the PB17 noise model is added before running the maps through the full analysis pipeline. These simulations are also used to generate an error term $\sigma_j$ for each observation, which is the standard deviation of all simulated angles for that observation. This construction includes the effect of sample variance, which is not relevant for our analysis because we make many observations of the same fixed CMB realization. The effect is negligible, however, because the angle error from sample variance alone is at least 16$\times$ smaller than the total statistical error for each observation.
The set of angles used in this analysis is shown in \figref~\ref{fig:angle_estimate}.

As in PB17, the absolute polarization angle of the instrument is calibrated using the $EB$-derived angle coadded across both seasons. This means that the weighted mean angle in this analysis is fixed to zero. While other non-$EB$ calibration sources exist, this procedure, described in \cite{Keating_2012}, is the most accurate way to calibrate the instrument. We show in \secref~\ref{sec:angle_subtraction} that this absolute polarization angle calibration has a negligible effect on our analysis.


The TOD filtering causes $E$-mode power to leak into $B$-mode maps, which can be estimated via simulation. 100 simulations of maps containing only CMB $C_\ell^{TT}$ and $C_\ell^{EE}$ power are run through the pipeline and used to construct $\hat{C}_{b,j}^{EB}$. A small amount of leakage is detected. It  is less than $6\%$ of the statistical error for any angle, resulting in a negligible ${<}0.01 ^\circ$ bias at any axion frequency probed in this analysis, and so we ignore its impact.

\subsection{Likelihood}
The noise between observations is assumed to be uncorrelated, and the errors on each angle Gaussian. The likelihood for the sinusoid parameters is then
\begin{equation} \label{eq:likelihood}
    \mathcal{L}(A, f, \theta) \propto e^{-\frac{1}{2} \chi^2(A, f, \theta)}
\end{equation}
where
\begin{equation} \label{eq:chisquare}
    \chi^2(A, f, \theta) = \sum_j^N \frac{(\langle s_j \rangle - \hat{\alpha}_j)^2}{\sigma_j^2}.
\end{equation}
The index $j$ represents a single observation, and the signal $\langle s_j \rangle$ is defined below. The likelihood normalization factor $\prod_j^N \sigma_j \sqrt{2 \pi}$ is dropped because we are only interested in relative likelihoods. Observations which have $\sigma_j$ at least two times larger than the mean value are discarded, since they contribute negligibly to the likelihood and are more likely to have systematic errors. This removed 23 observations, which constituted ${<}0.5\%$ of the statistical weight of the dataset. This leaves $N=515$ angles available for the analysis.

The signal $\langle s_j \rangle$ must be averaged over the duration of each observation:

\begin{align}
\begin{split}
    \langle s_j \rangle &= \frac{1}{t^{\text{end}}_j-t^{\text{start}}_j}\int_{t^{\text{start}}_j}^{t^{\text{end}}_j} dt\, A \sin{(2 \pi f t + \theta)} \\ 
    &= A \text{ sinc}(\pi f \Delta t_j) \sin(2 \pi f \bar{t}_j + \theta).  \label{eq:exact_signal}
\end{split}
\end{align}
$\Delta t_j \equiv t^{\text{end}}_j - t^{\text{start}}_j$ is the observation duration and $\overline{t}_j \equiv {(t_j^\text{end} + t_j^\text{start})}/{2}$ is the middle of the observation time. The effect of averaging is to wash out signals with oscillation periods that are shorter than the 4--8 hour observation periods.

\subsubsection{Frequency Domain} \label{sec:frequency_domain}
The frequency domain for our search is
\begin{equation} \label{eq:frequency_domain}
    0.02 \, \text{days}^{-1} \leq f \leq 0.45 \,\text{days}^{-1} .
\end{equation}
The dataset spans 694 days, 
so we are able to probe sinusoidal periods out to approximately that length. However, as explained in \secref~\ref{sec:systematics}, uncertainty in the relative HWP angle between the two seasons may show up as a low-frequency signal, leading us to set a minimum frequency of $0.02\,\text{days}^{-1}$. The maximum frequency is set to $0.45\,\text{days}^{-1}$ because the typical time-spacing between observations of a single patch is one day. This causes a significant loss in sensitivity around a frequency of $0.5\,\text{days}^{-1}$, as well as integer multiples of this frequency. Furthermore, at frequencies higher than $0.5\,\text{days}^{-1}$ the averaging effect in Eq. \eqref{eq:exact_signal} reduces the amplitude of the signal beyond the level of a few percent. For these reasons, we choose to only analyze frequencies up to $0.45\,\text{days}^{-1}$.

We assess the likelihood at a set of $1{,}492$ evenly-spaced frequencies. In the absence of noise, this allows the likelihood estimator to recover an injected signal at any frequency to within $5\%$ of the correct amplitude. This results in a frequency resolution about 5 times smaller than $1/694\,\text{days}^{-1}$, which is the spacing that a Discrete Fourier Transform of the same length of data would have. As a result, likelihood results from neighboring frequencies are highly correlated. This does not pose an issue for our analysis because we compare the results to simulated distributions which do not assume independence between different frequencies.

\subsubsection{Effect of Absolute Polarization Angle Calibration} \label{sec:angle_subtraction}
As described in \secref~\ref{sec:spectra_to_angles}, the absolute polarization angle of the instrument is calibrated by setting the weighted mean angle to zero. This means that our analysis is not sensitive to a static birefringence angle. To first order this is not an issue because we are searching for an oscillatory signal. However in the presence of such a signal, there will generally be some small shift in the mean angle caused by a partial oscillation. This shift is removed by the mean angle subtraction. This effect only becomes significant when there is ${\lesssim} 1$ oscillation period in the entire dataset. Our maximum period is 50 days, which corresponds to about 14 oscillation periods over the 694-day duration of our dataset. Therefore the subtraction in mean angle has negligible impact on the recovery of the signal amplitude.

\section{Null Tests} \label{sec:null_tests}
We run a series of null tests to check for systematic errors in our analysis before unblinding the data. The data are split along 15 possible sources of potential contamination, differenced to cancel any true signal, and tested for the presence of systematics. Twelve of the splits are the same as in PB17: ``moon distance'', ``sun distance'', ``rising vs. setting'', ``high vs. low PWV~(Precipitable Water Vapor)'', ``high vs. low elevation'', ``high vs. low gain'', ``sun above or below horizon'', ``1st vs. 2nd season'', ``1st vs. 2nd half of dataset'', ``left- vs. right-going scans'', ``left vs. right side of the focal plane'', and ``pixel polarization type''. We also add three new patch null tests, which pair two patches together in all three possible ways. Each split in these tests contains only the angles that were made from the $B$-mode maps from that patch, allowing us to probe for differences between patches.

\subsection{Methodology} \label{sec:null_analysis}
Two sets of angles are constructed for each test, one per split. The angles are constructed following the procedure in \secref~\ref{sec:analysis}, with the following modifications:
\begin{itemize}
    \item The $B$-mode maps contain only data from the split. This means that each map may contain less data or be eliminated altogether.
    \item For each split, a new map apodization mask is constructed per patch. These masks reflect differences in map coverage for each split. There are a few observations that have excessive noise in the unmasked region of some splits, and therefore cannot be null-validated. These observations' $B$-mode and $E$-mode maps are excluded from the analysis in this paper, as described in \secref~\ref{sec:mapmaking}. 
\end{itemize}

For each split, \textit{all} of the $E$-mode map data is used to construct $C_\ell^{EB}$, not just data from the split. This is because we wish to test for time-dependent systematic issues with the $B$-mode maps, and consider the coadded $E$-mode map validated in previous analyses (PB17).

A set of null statistics are computed for each test. For each of the two sets of angles `1' and `2', at each frequency $f$, we find the maximum likelihood estimation~(mle) of the amplitude $A^\text{mle}$ and phase $\theta^\text{mle}$ that maximizes $\mathcal{L}(A, f, \theta)$. The null statistic at a given frequency is
\begin{align}
    T_\text{null}(f) \equiv \frac{|A_{\rm null} (f)|^2}{\sigma \left( \Re(A_{\rm null}(f)) \right) ^2}
\end{align}
where
\begin{equation}
    A_\text{null}(f) \equiv (A^\text{mle}_1 e^{i \theta^\text{mle}_1} - A^\text{mle}_2 e^{i \theta^\text{mle}_2})(f).
\end{equation}
The $T_\text{null}$ can be thought of as a generalization of the difference of two discrete Fourier transforms: the correspondence would be exact if we had instantaneous observations evenly spaced in time and equally weighted. The normalization factor $\sigma \left( \Re(A_\text{null}) \right)$ is calculated using 500 simulations and ensures that $T_\text{null}(f)$ is distributed approximately as chi-squared with 2 degrees of freedom. This is because $\Re(A_\text{null})$ and $\Im(A_\text{null})$ are both distributed approximately as Gaussians with the same variance and zero mean. We do not rely on these analytic distributions since we compare the results to simulations, but we do rely on the normalization to allow for fair comparison across different tests and frequencies.

We calculate $T_\text{null}$ at $f=0$ for the 10 tests out of 15 that have associated periods within the frequency range of this search. This statistic is simply the difference in the weighted mean angles between the two splits. We include it because differences in mean angle between two splits could cause a spurious signal with a period associated with the split: for example, a difference in angle caused by the moon distance could generate a signal aligned with the ${\sim} 28$ days synodic cycle. Five tests are omitted from the $f=0$ analysis. ``1st vs. 2nd season'' and ``1st vs. 2nd half of dataset'' are excluded because the associated period is much longer than our maximum period of 50 days. The ``left- vs. right-going scans'', ``left vs. right side of the focal plane'', and ``pixel polarization type'' are excluded because both splits contain all observations, so a difference in mean angle would not generate a time-dependent signal. In total, there are 22390 $T_{\rm null}$ values: 1492 frequencies from \secref~\ref{sec:frequency_domain} for each of the 15 tests, and $f=0$ for 10 tests.

\subsection{Pass Criteria}
There are two pass criteria that need to be satisfied in order to pass the null tests. The first, ``pass criteria $\#1$,'' assesses failure in individual parts of the null test suite. The second, ``pass criteria $\#2$,''  uses Kolmogorov-Smirnov~(KS) tests to assess uniformity of all null suite probabilities.

Using $T_\text{null}$, five probability-to-exceed (PTE) values are computed. The lowest PTE among the five, $P_{\text{low}}^{\text{\#1}}$, is compared to a distribution generated from 500 simulations to calculate a global significance value, $\text{PTE}(P_{\text{low}}^{\text{\#1}})$. Pass criteria~$\#1$ requires more than $5\%$ of simulations to have a lower `lowest of five PTEs' value than $P_{\text{low}}^{\text{\#1}}$: i.e. $1-\text{PTE}(P_\text{low}^{\text{\#1}}) > 0.05$. 

The five PTE values are shown in \tabref~\ref{tab:PTE_table_1}, and are: 
\begin{enumerate}
    \item $\max_{t, f}  T_\text{null}$: The maximum $T_\text{null}$ value across all tests and non-zero frequencies. This tests for the presence of a systematic sinusoidal signal. 
    \item $\sum_{t, f} T_\text{null}$: The total chi-square value over all tests and non-zero frequencies, which assesses noise mis-estimation. 
    \item $\max_{t} \sum_f T_\text{null}$: The maximum per-test total chi-square value, which assesses issues with an individual test.
    \item $\text{max}_{f} \sum_t T_\text{null}$: The maximum per-frequency total chi-square value, which assesses issues with an individual frequency.
    \item $\text{max}_{t} T_\text{null}(f=0)$. The maximum difference in mean angle offset over the ten $f=0$ null tests.
\end{enumerate} 
The first four PTEs are calculated using only non-zero frequencies. 

\begin{table}[b]
\caption{\label{tab:PTE_table_1}%
The five null test PTE values used in the pass criteria~$\#1$.}
\begin{ruledtabular}
\begin{tabular}{cccc}
\textrm{PTE statistic}&
\textrm{Description}&
\textrm{PTE}\\
\colrule
$\max_{t, f} T_\text{null}$ & Spurious axion signal & 0.032\\
$\sum_{t, f} T_\text{null}$  & Total chi-square & 0.062\\
$\max_{t} \sum_f T_\text{null}$ & Bad test & 0.060\\ 
$\max_{f} \sum_t T_\text{null}$ &  Bad frequency & 0.246\\ 
$\max_{t} T_\text{null}(f=0)$ & Mean angle offset & 0.192\\
\end{tabular}
\end{ruledtabular}
\end{table}

Three PTE values are computed using KS tests, and pass criteria~$\#2$ is satisfied if more than $5\%$ of simulations have a lower value than $P_{\text{low}}^{\text{\#2}}$: i.e. $1-\text{PTE}(P_\text{low}^{\text{\#2}}) > 0.05$. 
Here $P_\text{low}^{\text{\#2}}$ is the lowest of the three PTEs.
These tests assess whether the probabilities of getting various $T_\text{null}$ values follow a uniform distribution. The KS test inputs and PTE results are shown in \tabref~\ref{tab:PTE_table_2}. The probability value returned by the KS test cannot be used to assess significance, since correlations between null tests result in a slightly non-uniform distribution of individual probabilities. Instead, the value of the KS statistic itself is compared to simulations.

\begin{table}[b]
\caption{\label{tab:PTE_table_2}
The three null test PTE values used in the pass criteria~$\#2$.}
\begin{ruledtabular}
\begin{tabular}{cccc}
\textrm{Axion KS Test inputs}&
\textrm{Description}&
\textrm{Number of inputs}&
\textrm{PTE}\\
\colrule
$\text{PTE}_{f, t} (T_\text{null})$ & Overall & 22380 & 0.128\\
$\text{PTE}_f ( \sum_t T_\text{null} )$ &  Per frequency & 1492 & 0.122\\
$\text{PTE}_t ( \sum_f T_\text{null} ) $ &  Per test & 15 & 0.190\\ 
\end{tabular}
\end{ruledtabular}
\end{table}

\subsection{Null test results}
For pass criteria~$\#1$, $1-\text{PTE}(P_\text{low}^{\text{\#1}}) = 0.124$.
For pass criteria~$\#2$, $1-\text{PTE}(P_\text{low}^{\text{\#2}}) = 0.148$. Both are greater than 0.05, and so the null tests pass.

While the null tests pass our stated criteria, all of the PTE values are relatively low. This is potentially indicative of a systematic issue, which we investigated. The low PTEs do not appear to be caused by particular null splits or frequency ranges. Two of the PTEs in \tabref~\ref{tab:PTE_table_1} (`bad test' and `bad frequency') explicitly look for the worst test and frequency, and were not discrepant enough to fail pass criteria~$\#1$. Pass criteria~$\#2$ is sensitive to non-uniformities in the $T_{\rm null}$ distribution and also passed. In addition to these summary statistics, visual inspection of the full set of $T_{\rm null}$ values plotted for each test and frequency do not reveal large asymmetries between different null splits and frequency ranges. A set of correlated, low PTEs is not surprising given a high total chi-square value. The real data has a total chi-square value $20 \%$ larger than the simulation average. This is larger than all but $6 \%$ of simulations as shown in \tabref~\ref{tab:PTE_table_1}. Simulations with such a large total chi-square value have the overwhelming majority of their PTE values less than $0.2$.

A high total chi-square value could indicate a problem with the noise model. The noise model is based on Monte-Carlo simulations of white noise timestreams, as described in PB17. It was validated on the full coadded spectra, so we do not expect it to be a source of error. However, to check for noise mis-estimation, we computed a set of angles using a different 100 realizations of signflip noise $B$-mode maps, which are created by randomly reversing the sign of the maps made for each 15~minute CES. This cancels out the true signal but maintains the noise properties of the real data. We compared $\sigma \left(\Re(A_\text{null}) \right)$ from signflip simulations to those computed using Monte-Carlo noise simulations, and found they agreed to within $5\%$, and in the wrong direction to alleviate the total chi-square tension. Having found no obvious discrepancy under a different noise model, and with the null tests having passed our pre-determined criteria, we decided that a somewhat elevated total chi-square value would not prevent us from unblinding the data.

\section{Systematic Errors} \label{sec:systematics}

In this section we examine the impact of several sources of systematic error, including uncertainty in the position of the HWP, any differences in average angle between the three observing patches, and foregrounds. Although we rely on some of the systematic pipeline used in PB17, our analysis is fundamentally different because we are looking for \textit{time-dependent} sources of error. Systematics that provide a constant $C_\ell^{EB}$ offset are irrelevant. Therefore many of the PB17 systematics are of no concern, while two others, the HWP and differences between patches, require careful consideration.
The typical maximum likelihood sinusoid amplitude generated by statistical noise alone is $0.26^\circ$.
Therefore any oscillatory systematic that generates an amplitude much less than $0.26^\circ$ should have negligible impact on the limits in this work.

\subsection{HWP Position Uncertainty} \label{sec:hwp_systematic}
As discussed in \secref~\ref{sec:telescope}, angle errors that are introduced by the stepped rotation of the HWP during the first observing season are a significant source of time-dependent uncertainty. The HWP angle was changed about 60 times during the first half of the first season, 4 times during the second half of the first season, and was then fixed for the second season. During each rotation the HWP was commanded to rotate in increments of $11.25^\circ$. However, when examining the polarization angles derived from an alternative calibration source, the Crab Nebula (Tau A), it became clear that the angles during the first season exhibited larger variance than during the second season. Furthermore, there was an offset in angle between the two seasons. Both phenomena could be explained if the HWP did not step exactly to the commanded position each time. While the exact HWP-induced offset at each step is unknown due to statistical error on the Tau A measurements, PB17 found that the discrepancy could be explained by adding a systematic error of $0.56^\circ$ for each step, corresponding to a typical HWP offset of $0.28^\circ$, in quadrature with the statistical error.

This systematic is highly important for this analysis because it corresponds to a unknown time-dependent shift in the instrument polarization angle, which could mimic an axion-generated birefringence signal. The HWP-induced offset between the first and second seasons, for example, is degenerate with a signal with a period of one year. To assess the impact at higher frequencies, we ran 500 simulations where a random angle offset, drawn from a Gaussian distribution with $\sigma = 0.56^\circ$, was assigned to each HWP angle. Then we generated timestreams where each observation's CMB angle corresponded to the offset given by the HWP angle. The maximum likelihood sinusoid amplitude was then estimated at each axion frequency.

\begin{figure}
    \includegraphics[width=0.48\textwidth]{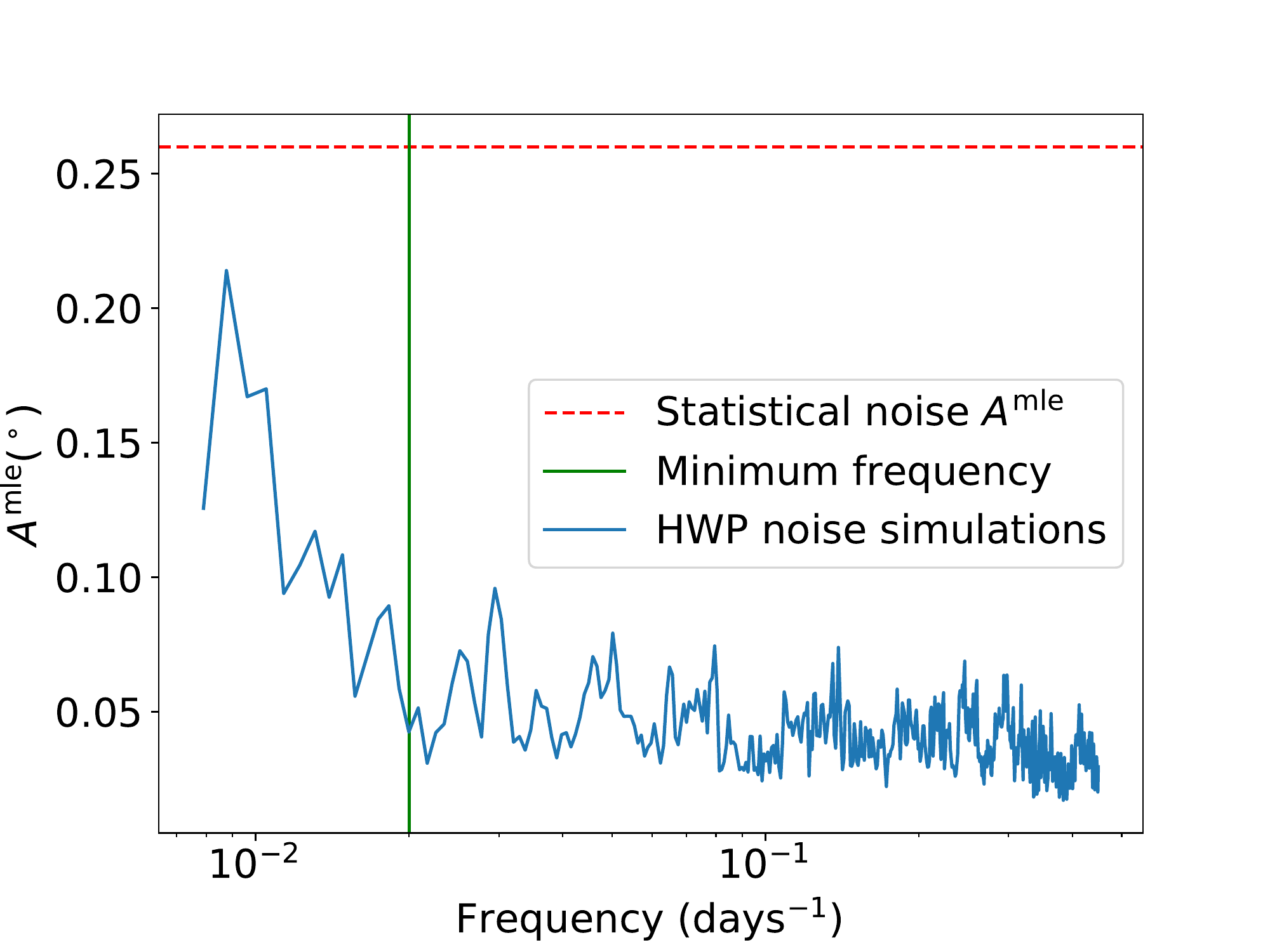}
    \caption{The estimated typical maximum likelihood amplitude signal of the HWP angle offsets. 500 simulations containing only random offset angles were generated and run through the likelihood. The median recovered amplitude is shown here. The HWP step cadence begins to generate larger systematic issues at periods ${>}50$ days. Based on these results, the maximum period used in this analysis is $50$ days.}
    \label{fig:hwp_jiter}
\end{figure}

The result is shown in \figref~\ref{fig:hwp_jiter}. At high frequencies the HWP rotation cadence doesn't correspond to any specific frequency. Since observations have a statistical error of at least $2^\circ$ and the typical HWP offset is $0.56^\circ$, the impact at high frequencies is minimal. On average these offsets cause only a $1-2\%$ increase in the maximum likelihood amplitude estimate at a given frequency relative to statistical noise alone, and so we ignore the impact of this HWP offset noise in our likelihood. At low frequencies, however, the last few HWP steps in the first season and the offset between the first and second seasons begin to translate into larger sinusoidal signals. We emphasize that because the exact HWP-induced offset at each step is unknown, the numbers shown in \figref~\ref{fig:hwp_jiter} are only indicative of the average effect we expect, not the true effect present in our data. Because the HWP step cadence begins to generate larger systematic issues at periods ${>}50$ days, there is a higher potential to generate a signal that would cause false detection. Based on these results, we chose to set 50 days as the longest period we analyze.

It is possible that the HWP offsets follow a more pathological model than what we assume and mimic a sinusoidal signal at some frequency in our domain. The ``1st half vs 2nd half'' null test described in \secref~\ref{sec:null_tests} tests for this,
because the time periods align with when the HWP was stepped vs. fixed. As stated in that section, there is no signal observed in this null test.

\subsection{Differences in Patch Mean Angle} \label{sec:Patch mean angle}

Since a typical day involves observing all three patches sequentially, any differences in the mean angle of each of the three patches could generate a oscillatory signal with a period of one day. However, as explained in \secref~\ref{sec:frequency_domain}, our upper frequency bound is $0.45\,\text{days}^{-1}$ and so we are not sensitive to this signal. However, variations in the observing sequence can generate signals at other frequencies, and simulations of patch mean angle offsets reveal a transfer of ${\leq} 12\%$ of any offset into any frequency in domain we do observe. Fortunately, our null tests directly calculate the difference in mean angle between patches. Since they are consistent with simulations, which do not contain foreground power, we do not see evidence for any foreground or systematic patch mean angle differences.

One known effect that will cause offsets between the patch mean angles is sample variance, because each patch has an independent sample-variance-induced $C_{\ell}^{EB}$ signal. This effect is included in the $\Lambda {\rm CDM}$ simulations, but if it were large we would need to correct for its effects. It can be calculated both analytically and through simulations, and in each patch, the error from sample variance is $\sigma \approx 0.12^\circ$. This is 3$\times$ smaller than the statistical error alone on each patch mean angle. Furthermore, since the patch mean angle offsets transfer ${\leq} 12\%$ into any given frequency, the typical $A^\text{mle}$ generated by cosmic variance is a negligible ${\lesssim} 0.01^\circ$.

\subsection{Foregrounds and Other Instrumental Systematics} \label{sec:foregrounds_other}
Any $C_{\ell}^{EB}$ contribution from dust and synchrotron radiation should be time-independent and simply contribute to the patch mean angle differences. These differences
were addressed in \secref~\ref{sec:Patch mean angle}. Point sources may be variable in time, but the strongest are masked as discussed in PB17.

PB17 also makes many other estimates related to the calibration, analysis effects, and known instrumental systematics. We expect the multiplicative effects to have a similar impact on the $C_{\ell}^{EB}$ spectrum as they do on the $C_{\ell}^{BB}$ spectrum, which PB17 determined was $6\%$. Since the angle estimates are linear in $C_{\ell}^{EB}$, we would then roughly expect a overall $6\%$ uncertainty on amplitude estimation. Since this is relatively small and does not bias our results, this is negligible uncertainty in our reported results.

The other instrumental systematics deal with effects that could potentially cause spurious additive $C_{\ell}^{EB}$, including differential gain, gain drift, differential beam size, differential ellipticity, differential pointing, boresight pointing, and electrical crosstalk. These can impact our analysis if they generate a time-dependent $C_{\ell}^{EB}$. An analysis of $C_{\ell}^{EB}$ bias was done in PB14, which noted that all sources provided ${\lesssim} 0.02^\circ$ bias in the coadded spectra. Using a combination of analytic estimates \cite{Shimon_2008} and simulations, we modified the analysis to account for per-observation $C_{\ell}^{EB}$ and determined that all such sources were either time-independent or provided a negligibly small time-dependent contribution as compared to the statistical noise.

\section{Results} \label{sec:results}
We search for the presence of a sinusoidal signal, and finding no statistically significant indication, place appropriate bounds over the frequency range of the search. By accounting for the axion field amplitude at the telescope, these translate to bounds on the axion-photon coupling over a range of axion masses.

\subsection{Search for a Signal}
In order to detect a signal, we form a test statistic $\Delta \chi^2$ which is large in the presence of a sinusoidal signal above the background. For each frequency in the discrete frequency domain, the MLE phase and amplitude are found, and the difference relative to the $\chi^2$ with no signal is calculated: 
\begin{equation}
    \Delta \chi^2(f) \equiv \chi^2(A=0) - \chi^2(A^\text{mle}(f), f, \theta^\text{mle}(f)).
\end{equation}
The test statistic is the largest $\Delta \chi^2(f)$ among all frequencies:
\begin{equation}
    \Delta \chi^2 \equiv \text{max}_f (\Delta \chi^2(f)) .
\end{equation}
  
This statistic is optimal in the Neyman-Pearson sense in that it maximizes the probability of rejecting the no-signal hypothesis if the alternative hypothesis of a sinusoidal signal is true \cite{PDG:2022}. The distribution of the statistic is approximately chi-squared, but we do not rely on this, instead comparing to a distribution of test statistics computed from 500 $\Lambda {\rm CDM}$ + noise simulations. The p-value is the fraction of background simulation test statistic values that exceed that of the real data.

\begin{figure}
    \includegraphics[width=0.5\textwidth]{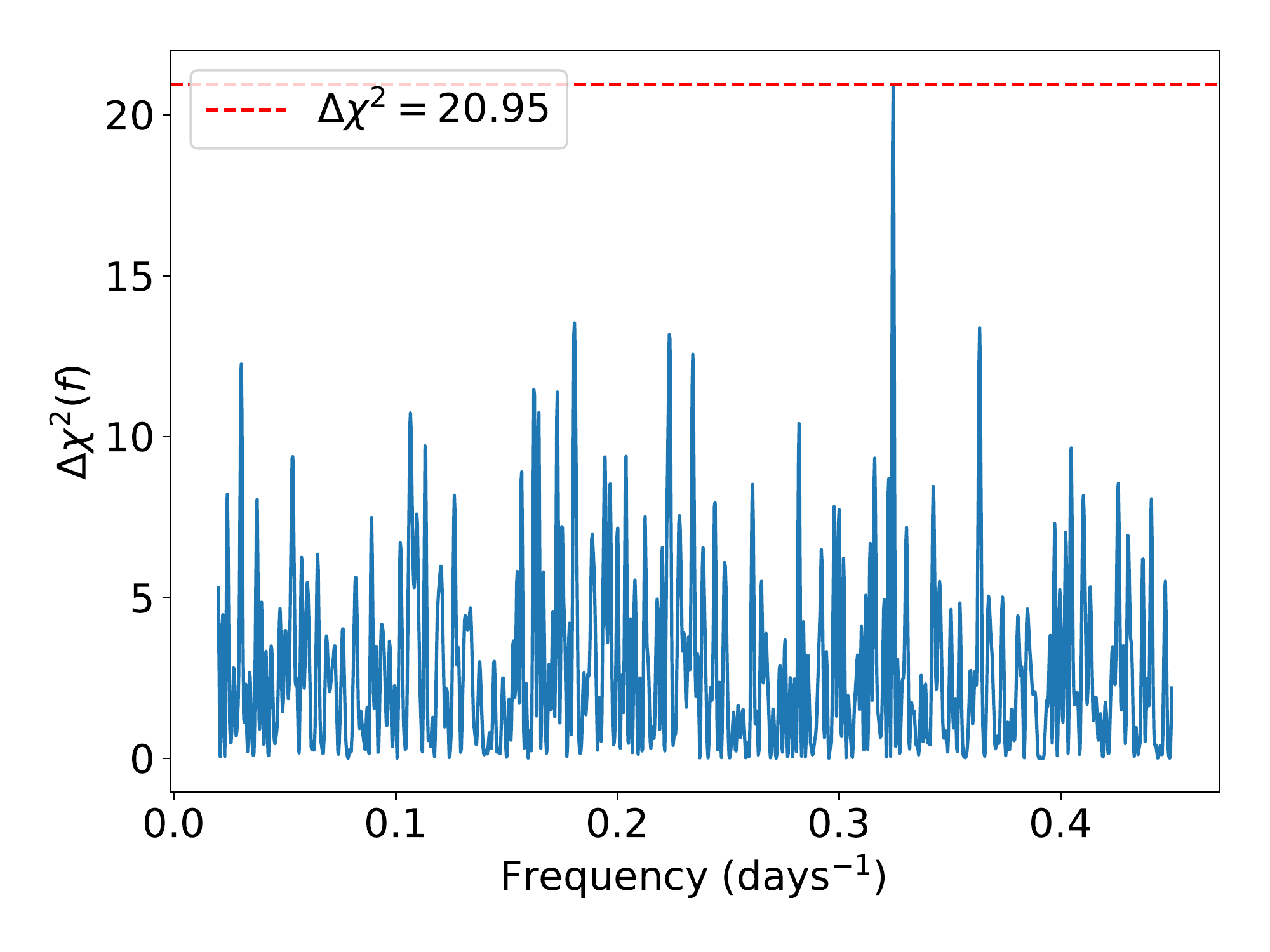}
    \includegraphics[width=0.5\textwidth]{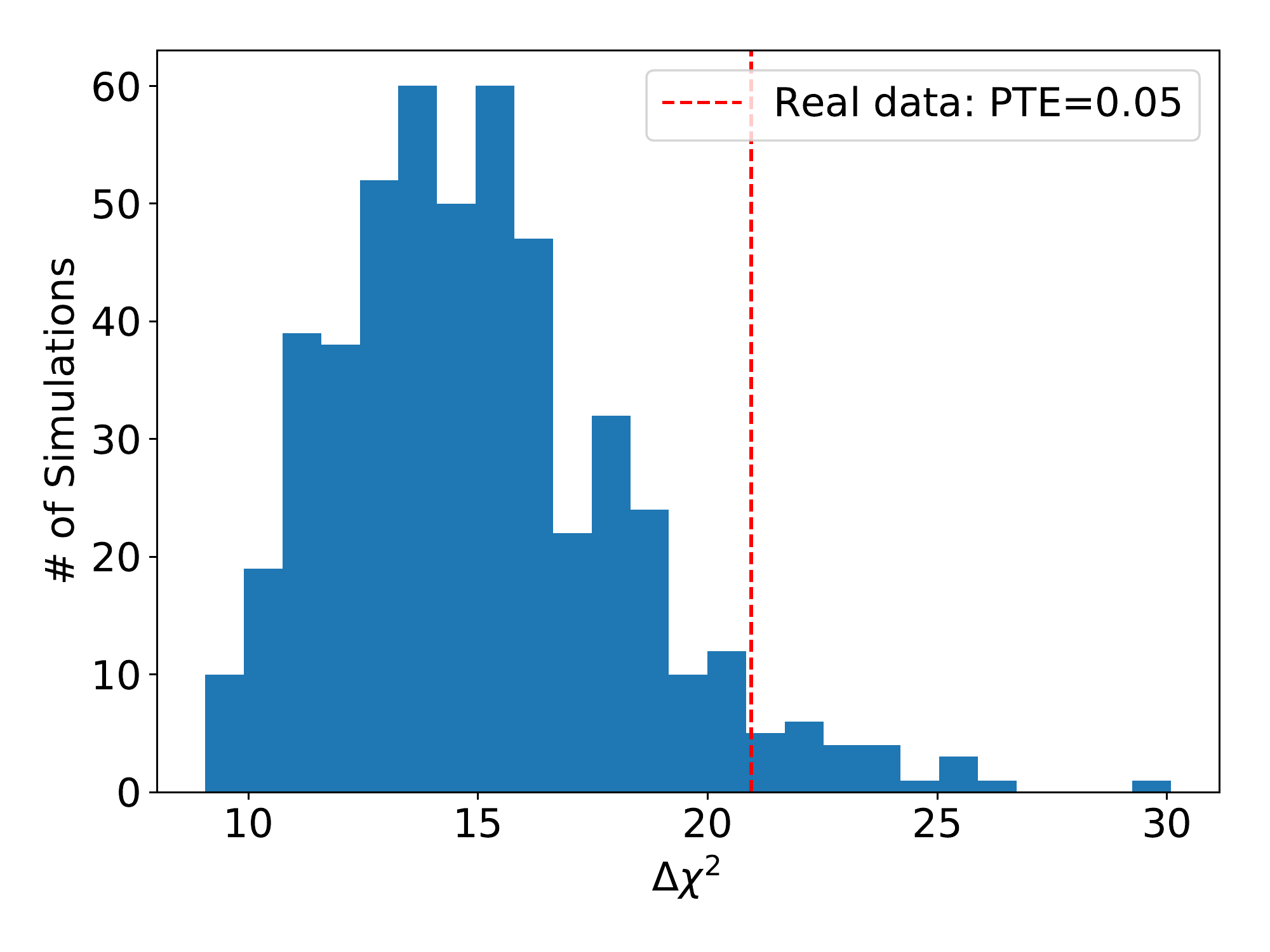}
    \caption{The significance test for the presence of an axion signal.
      \textit{Top}: The calculation of $\Delta \chi^2$. For each frequency in the discrete frequency domain, the MLE phase and amplitude are found, and $\Delta \chi^2(f)$ is calculated.
      $\Delta \chi^2$ is the largest among these.
      \textit{Bottom}: The significance of $\Delta \chi^2$ as compared to a set of 500 simulations.}
    \label{fig:significance_test}
\end{figure}

The selection of $\Delta \chi^2$ and its location relative to the distribution is shown in \figref~\ref{fig:significance_test}. The real data has a p-value of 0.048, corresponding to $1.7 \sigma$ significance. While this p-value is relatively small, it is not significant enough to claim detection of a signal above the noise background. Future analysis of the remaining three seasons of {\sc Polarbear} will aid in determining if this result is simply a statistical fluctuation. 

When assessing the significance of a result with many trials, in this case discrete frequencies, it is often useful to report the experimental sensitivity. This is defined as the signal amplitude which would produce a detection at some significance level. We can approximate the $\Delta \chi^2$ required for any significance level by applying a fit to the distribution shown in the bottom of \figref~\ref{fig:significance_test}. A $3 \sigma$ detection (PTE = 0.0013) would require $\Delta \chi^2=28.5$, yielding an experimental sensitivity of $A(3\sigma) = 1.15 ^\circ$. The $5 \sigma$ (PTE $= 2.8 \times 10^{-7}$) experimental sensitivity is $A(5\sigma) = 1.43 ^\circ$.

\subsection{Upper Limit} \label{sec:upper limit}
\begin{figure}[htbp]
    \includegraphics[width=0.48\textwidth]{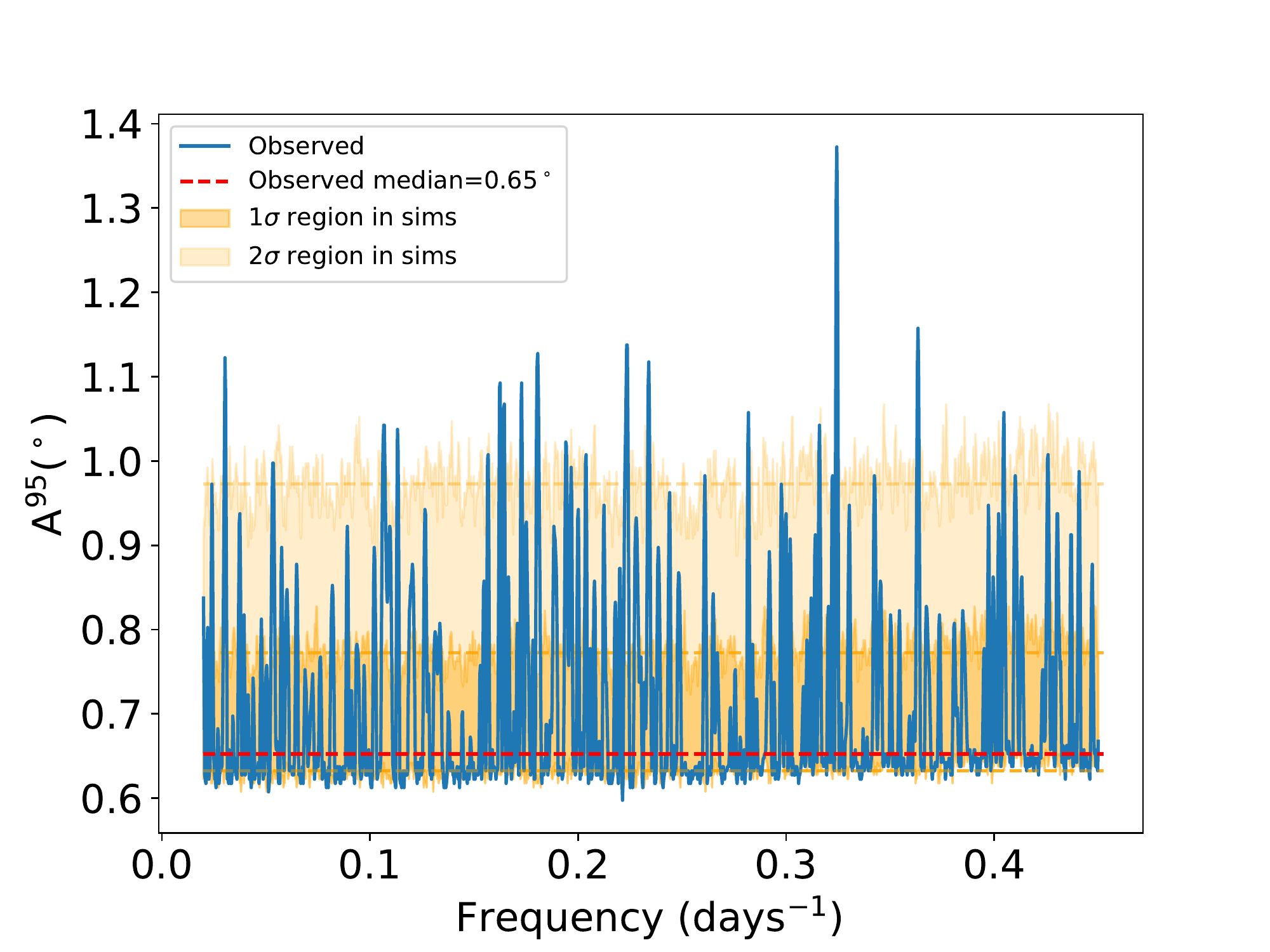}
    \caption{A $95\%$ upper confidence limit on the presence of a sinusoidal signal at each frequency. The $1 \sigma$ and $2 \sigma$ regions contain approximately $68\%$ and $95\%$ of the upper bounds calculated from 500 simulations at each frequency. The median values for these regions are indicated with dashed lines. Many individual frequencies exceed the $2 \sigma$ threshold, which is to be expected based on the large number of test frequencies, and does not necessarily indicate a detection.}
    \label{fig:upper_limit}
\end{figure}

As we do not detect a globally significant axion-like oscillation, at each frequency we compute a frequentist upper limit on the signal amplitude at $95\%$ confidence level. The likelihood for the amplitude as a function of frequency is 
\begin{align}
    \mathcal{L}_{A}(A, f) = \int_0^{2 \pi} d \theta \mathcal{L}(A, f, \theta) P(\theta),
\end{align}
with a uniform probability distribution for the phase
\begin{equation} \label{eq:Ptheta}
    P(\theta) = \frac{1}{2 \pi},\; 0 \leq \theta < 2 \pi.
\end{equation}
We construct confidence intervals following the Neyman procedure \cite{PDG:2022} with the following ordering: intervals [0, $A^{95}(f)$] are defined with an upper limit $A^{95}(f)$ at each frequency $f$ such that
\begin{equation} \label{eq:confidence_limits}
    P(A^{\rm mle} \leq A^{\rm mle}_{\rm obs} | A^{95})(f) = 0.05.
\end{equation}
Here $A^{\rm mle}_{\rm obs}$ is the observed maximum likelihood amplitude calculated from the data using $\mathcal{L}_{A}$. $P(A^{\rm mle}| A)(f)$ is the probability of calculating $A^{\rm mle}$ given a true signal amplitude $A$ plus background noise. Therefore $P(A_{\rm mle} \leq A^{\rm mle}_{\rm obs} | A)(f)$ represents the probability that, given a true signal of amplitude $A$, $A^{\rm mle}$ would be less than or equal to the value observed. This probability is a monotonically decreasing function of $A$, and so it is less than $0.05$ for all $A$-values above the upper limit. In other words, signal amplitudes that would generate an $A^{\rm mle}$ as low or lower less than $5\%$ of the time are excluded from the confidence interval. The limits calculated by this procedure are shown in \figref~\ref{fig:upper_limit}. The median upper bound is $0.65^\circ$. The bound varies over frequencies as expected, in analogy to the typical behavior of a Fourier transform.

The probability distribution $P(A^{\rm mle}| A)(f)$ is generated by calculating $A^{\rm mle}$ from 500 simulated angle timestreams of background noise with an injected signal of amplitude $A$, frequency $f$, and random phase between $0$ and $2\pi$. This is done for all frequencies and discretized array of $A$-values, which are smoothed to create a continuous probability distribution. The results are checked for convergence of the median upper limit to the ${<}1\%$ level. $P(A^{\rm mle}| A)(f)$ is almost the same for each frequency. The primary difference comes from the finite duration of each observation, which reduces the strength of the bounds at higher frequencies according to the sinc function in Eq.~\eqref{eq:exact_signal}. We can approximate the impact of this by calculating $\text{sinc}(\pi f \overline{\Delta t})$, where $\overline{\Delta t} = 6.4$ hours is the weighted mean observation duration. The maximum amplitude reduction occurs at the largest frequency and is ${<}2.5\%$. While this effect is included in the constraint we place, it is small enough that we still report the median bound over the full frequency range.

\subsection{Constraints}\label{sec:constraints}
The median angle bound $A^{95}$ represents a search for a sinusoidal signal in the CMB data independent of any axion model. Translating this into a constraint on the axion-photon coupling constant $g_{\phi \gamma}$ requires specifying model parameters. In F19, BK22, and SPT22, this was done by simply requiring that the local axion energy density equal the average Milky Way dark matter energy density, which we will call the `deterministic' case, following reference \cite{Centers_2021}. This means the axion field has amplitude $\phi_{\rm DM}$, where $\phi_{\text{DM}}^2 m_\phi^2 /2 = \rho_0$ and $\rho_0$ is the local density of dark matter. This gives the relation
\begin{align}
  g_{\phi \gamma} &= (1.6 \times 10^{-11} \, \text{GeV}^{-1}) \nonumber \\
  &\quad\times \left( \frac{A}{1^\circ} \right)
  \times \left( \frac{m_\phi}{10^{-21} \, \text{eV}} \right)
  \times \left( \frac{\kappa \times \rho_0}{0.3 \,\text{GeV}/\text{cm}^3} \right)^{-{1}/{2}} .
\end{align}
Here $A$ is the rotation amplitude, $m_\phi$ is the axion mass, and $\kappa$ is the fraction of the dark matter that the axion constitutes. Using $\kappa=1$, $\rho_0 = 0.3 \, \text{GeV}/\text{cm}^3$, and $A^{95}=0.65^\circ$, this constraint is 
\begin{equation} \label{eq:deterministic_constraint}
    g_{\phi \gamma} < (1.1 \times 10^{-11} \, \text{GeV}^{-1}) \times \left( \frac{m_\phi}{10^{-21} \text{eV}}\right).
\end{equation}

Reference \cite{Centers_2021}, however, has recently pointed out that this constraint is inappropriate when the timescale of the experiment is much less than the axion field coherence time, as it is in our case. When considering the local axion field as the sum of many individual wave modes, each with random phase, the amplitude $\phi_0$ at any given time is a Rayleigh-distributed stochastic variable centered on $\phi_\text{DM}$ that varies with time on the coherence timescale $\tau_\text{coherence} \sim (f {v_\text{MW}^2}/{c^2})^{-1}$, where the virial velocity is $v_\text{MW} \approx 10^{-3} c$. We will call this the `stochastic' case. This random phase model has been shown to roughly agree with simulations of fuzzy dark matter \cite{Hui_2017}. The largest frequency we consider in our analysis is $f_{\max} = 0.45 \, \text{days}^{-1}$, yielding a minimum coherence time of about 6000 years: therefore, for the purposes of our experiment we can approximate $\phi_0$ as a fixed random variable. The phase is treated in the same manner with a uniform probability distribution. To form the likelihood for $g_{\phi \gamma}$ we integrate over these two parameters:

\begin{widetext}
\begin{equation} \label{eq: likelihood_g}
    \mathcal{L}_{g_{\phi \gamma}}(g_{\phi \gamma}, f) = \int_0^{\infty} d \phi_0 \int_0^{2 \pi} d \theta\, \mathcal{L}({g_{\phi \gamma} \phi_0}/{2}, f, \theta) P(\phi_0) P(\theta) .
\end{equation}
\end{widetext}
Here $P(\phi_0)$ is a Rayleigh distribution centered on $\phi_\text{DM}$:
\begin{equation} \label{eq:Pphi}
    P(\phi_0) = \frac{2 \phi_0}{\phi_\text{DM}^2} e^{-\frac{\phi_0^2}{\phi_\text{DM}^2}}.
\end{equation}

The resulting median $95\%$ upper limit in this stochastic case is calculated in the same manner as \secref~\ref{sec:upper limit}, and is
\begin{equation} \label{eq:stochastic_constraint}
    g_{\phi \gamma} < (2.4 \times 10^{-11} \,\text{GeV}^{-1}) \times \left( \frac{m_\phi}{10^{-21} \,\text{eV}} \right).
\end{equation}
This is $2.2 \times$ larger than the deterministic case.\footnote{Reference [22] reports an increase of $2.7 \times$ when using the frequentist approach. The difference possibly stems from the use of a different likelihood (they assume that the data is uniformly spaced in time) and/or differences between individual realizations of the noise. When comparing the Bayesian deterministic approach and the Bayesian stochastic approach with a uniform amplitude prior we see a $10 \times$ increase, in agreement with [22].} The increase is due to the possibility that we happen to be observing at an unlucky time when $\phi_0 < \phi_{\text{DM}}$, which generates a smaller signal amplitude at a given $g_{\phi \gamma}$.

We briefly comment on the choice between frequentist and Bayesian statistics in this analysis. In the frequentist approach adopted here, we generate confidence intervals according to Eq.~\eqref{eq:confidence_limits}. In a Bayesian approach, we would instead generate a posterior probability distribution for $g_{\phi \gamma}$, which requires choosing a prior $P(g_{\phi \gamma})$. 

In the deterministic case, or equivalently when placing limits on the sinusoid amplitude $A$, the results roughly agree for several choices of prior. With a uniform prior on $A$, the median Bayesian upper limit is $0.60^\circ$. If we had instead chosen to parameterize the signal (Eq. \eqref{signal}) as 
\begin{equation} \label{eq:sinecosine}
    \beta_{\text{CMB}}(t) = B\sin(2 \pi f t) + C\cos(2 \pi f t),
\end{equation}
applying uniform priors on $B$ and $C$ give a median upper limit of $0.71^\circ$. 
Both of these results are close to the median frequentist $0.65^\circ$ upper limit we report in \secref~\ref{sec:upper limit}.  

In the stochastic case there is a large dependence on the choice of prior and parameterization. We must integrate over $P(\phi_0)$, and the non-zero probability of small $\phi_0<\phi_{\text{DM}}$ values means that the posterior for $g_{\phi \gamma}$ has a long tail at large $g_{\phi \gamma}$. A uniform prior on $g_{\phi \gamma}$ results in a Bayesian upper limit $10 \times$ greater than the deterministic case, much larger than the frequentist result. Futhermore, applying uniform priors in the sine-cosine parameterization (Eq. \eqref{eq:sinecosine}) causes the upper bound to diverge. To resolve these issues, reference \cite{Centers_2021} advocates choosing a Berger-Bernardo prior for $g_{\phi \gamma}$, which like the Jeffreys's prior is parameterization-invariant, and in their analysis agrees with the frequentist result to within $3\%$. Unfortunately, for our likelihood we found no simple analytic form for this prior and it is difficult to accurately compute numerically. Due to issues with defining a Bayesian prior in the stochastic case, we choose to report a frequentist limit.

\begin{figure*}
  \includegraphics[width=0.95\textwidth]{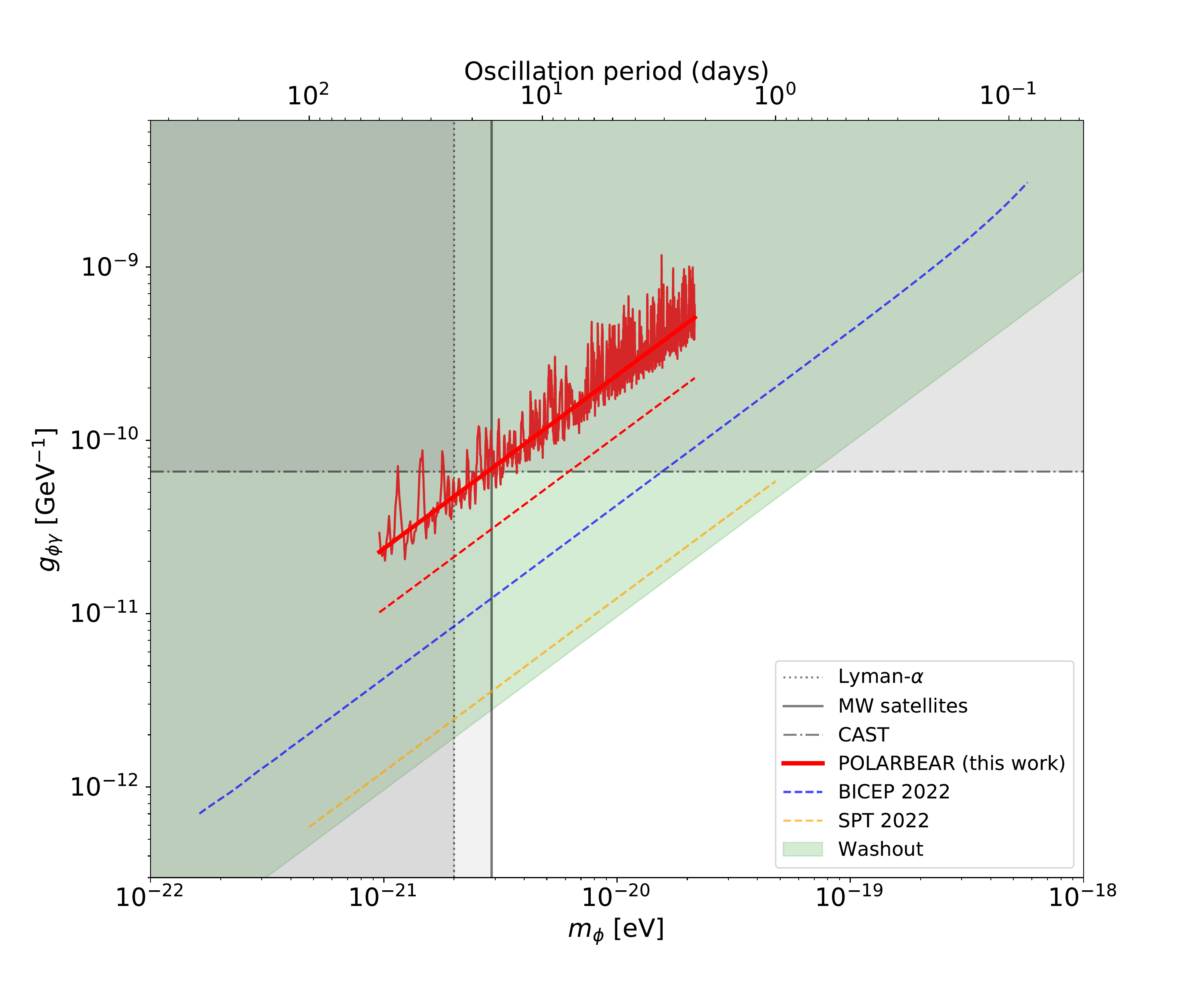}
  \caption{Bounds on the axion-photon coupling as a function of axion mass. Shown in solid red is the $95\%$ upper limit obtained in this work using a stochastic local axion field amplitude and assuming the axion constitutes all the dark matter (see \secref~\ref{sec:results}).
    The lighter oscillating result is the exact bound, and the median result,
    $g_{\phi \gamma} < (2.4 \times 10^{-11} \, \text{GeV}^{-1} ) \times ({m_\phi}/{10^{-21} \, \text{eV}})$, is shown in darker red. The median deterministic bounds for {\sc Polarbear}, BICEP, and SPT are the dashed lines: these are the published results in \cite{BK_2022} and \cite{SPT_2022}. The green `washout' bound was calculated in F19 from the lack of CMB polarization suppression. Lower bounds on the axion mass come from Milky Way satellites \cite{DES:2020fxi} and the Lyman-$\alpha$ forest \cite{Ir_i__2017}.
    Upper bounds on the axion-photon coupling from CAST \cite{CAST_2017} are also shown. 
    }
    \label{fig:coupling_constraints}
\end{figure*}

This limit is shown in \figref~\ref{fig:coupling_constraints}, along with a selection of other constraints. Our primary result is the constraint on $g_{\phi \gamma}$ assuming that the axion field amplitude is a Rayleigh-distributed stochastic variable. It is shown in full detail along with the median limit from Eq.~\eqref{eq:stochastic_constraint}. The median deterministic constraint, Eq.~\eqref{eq:deterministic_constraint}, is also shown. The published results from BK22 and SPT22 are deterministic Bayesian upper bounds with a uniform prior on $P(g_{\phi \gamma})$. None of the other bounds shown in \figref~\ref{fig:coupling_constraints} come from assuming a value for the axion field in a dark matter halo.  We emphasize that this is the first CMB analysis of this effect that we are aware of to include the local stochastic nature of $\phi_0$.

In both the stochastic and deterministic cases, the bounds apply over the frequency range presented in Eq.~\eqref{eq:frequency_domain},
which corresponds to the axion mass range
\begin{equation} \label{eq:mass_domain}
    9.6 \times 10^{-22} \, \text{eV} \leq m_\phi \leq 2.2\times 10^{-20} \,\text{eV}.
\end{equation}

\section{Conclusion} \label{sec:conclusion}
We have used {\sc Polarbear}  data to search for a coherent, all-sky, sinusoidal oscillation of the CMB polarization angle in time. We do not detect such a signal, and place a median $95 \%$ upper limit of $0.65^\circ$ on the sinusoid amplitude over oscillation frequencies between $0.02 \, \text{days}^{-1}$ and $0.45 \, \text{days}^{-1}$. We use these results to constrain the coupling between electromagnetism and an axion, here defined as an ultralight pseudoscalar field, under the assumption that the axion constitutes all of the dark matter. The signal depends on the value of the axion field at the telescope, and under the assumption that the field amplitude is a Rayleigh-distributed stochastic variable,
we set the limit $g_{\phi \gamma} < (2.4 \times 10^{-11} \, \text{GeV}^{-1}) \times ({m_\phi}/{10^{-21}\,\text{eV}})$
over the mass range $9.6 \times 10^{-22} \, \text{eV} \leq m_\phi \leq 2.2\times 10^{-20} \,\text{eV}$.

Three additional seasons of {\sc Polarbear} data have been collected in addition to the two seasons analyzed here, and we anticipate that analyzing them will improve our data volume by a factor of ${\approx} 2\text{--}3$,
with a corresponding ${\approx} 60\%$ improvement
in the constraints. This data will also possibly allow us to probe lower frequencies (smaller axion masses), because these seasons are not affected by the HWP position uncertainty that restricted the frequency range in this analysis. A promising avenue for placing constraints several times better than this with {\sc Polarbear} data lies with measurements of Tau A, which was used as a polarization calibration source, and was precisely measured during the five observing seasons between 2012 and 2016. These measurements present a different challenge than the CMB because the axion field at the source needs to be carefully considered, whereas in our analysis the $\mathcal{O}(10^5)$ year duration of recombination allowed for ignoring the source term. Nonetheless, this additional data provides another avenue to search for the presence of axions using {\sc Polarbear} data.

Several future CMB experiments, including the Simons Array \cite{Suzuki_2016}, Simons Observatory \cite{Ade_2019}, and CMB-S4 \cite{CMBS4}, should be able to perform a similar analysis with improved constraints. This analysis imposes no additional requirements on the design or operation of these experiments, it simply requires making many measurements of the CMB over an extended period of time. The sensitivity is not fundamentally limited by anything other than the precision of the polarization measurements, unlike the CMB washout effect, which is limited by cosmic variance \cite{Fedderke_2019}. The unknown amplitude of the axion field at the telescope is the chief source of model-dependence, but it is well described by a probability distribution that can be treated statistically when placing bounds. Unlike many other astrophysical measurements, this analysis does not suffer from significant modelling uncertainty at the polarization source due to the well-understood nature of the CMB. In the rapidly growing field of axion searches, this should allow future CMB experiments to provide increasingly competitive measurements of the axion-photon coupling constant.

\begin{acknowledgments}
The {\sc Polarbear} project is funded by the National Science Foundation under grants AST-0618398 and AST-1212230. The analysis presented here was also sup- ported by Moore Foundation grant number 4633, the Simons Foundation grant number 034079, and the Templeton Foundation grant number 58724. C.B. acknowledges support from the COSMOS project of the Italian Space Agency (\href{https://www.cosmosnet.it}{cosmosnet.it}), and the INDARK Initiative of the INFN (\href{http://web.infn.it/CSN4/IS/Linea5/InDark}{web.infn.it/CSN4/IS/Linea5/InDark}). The James Ax Observatory operates in the Parque Astron\'omico Atacama in Northern Chile under the auspices of the Comisi\'on Nacional de Investigaci\'on Cient\'ifica y Tecnol\'ogica de Chile (CONICYT). Y.C. acknowledges the support from JSPS KAKENHI grant Nos. 18K13558 and 21K03585. J.E. acknowledges funding from the European Research Council (ERC) under the European Union’s Horizon 2020 research and innovation program (Grant agreement No. 101044073). G. F. acknowledges the support of the European Research Council under the Marie Sk\l{}odowska Curie actions through the Individual Global Fellowship No. 892401 PiCOGAMBAS. This work was supported by World Premier International Research Center Initiative (WPI), MEXT, Japan. The work of MH was supported by JSPS KAKENHI Grant Number JP22H04945. In Japan, this work was supported by JSPS KAKENHI grant Nos. 18H05539 and 19H00674.  This work was supported in part by JSPS core-to-core program number JPJSCCA20200003.  Work at LBNL is supported in part by the U.S. Department of Energy, Office of Science, Office of High Energy Physics, under contract No. DE-AC02-05CH11231. HN acknowledges the support from the JSPS KAKENHI Grant Number JP26800125. CR acknowledges support from the Australian Research Council's Discovery Projects scheme (DP210102386). ST acknowledges JSPS Overseas Research Fellowships and JSPS KAKENHI Grant No. JP14J01662. Support from the James B. Ax Family Foundation is acknowledged.

\end{acknowledgments}

\bibliography{refs}

\end{document}